\documentclass[12pt,titlepage]{article}
\usepackage[pdftex]{graphicx}
\usepackage[numbers]{natbib}

\begin{document}

\title{\Huge Great Surveys of the Universe}

\author{{\LARGE Steven T.~Myers} \\ National Radio Astronomy Observatory \\ 
  P.O.Box O, Socorro, NM, 87801 \\ 
  {\small contact author: {\tt smyers@nrao.edu}}
}

\date{Astro2010 State of the Profession position paper \\[3ex]
\begin{minipage}[h]{6.5in}
  \begin{center}
  \includegraphics[width=6in]{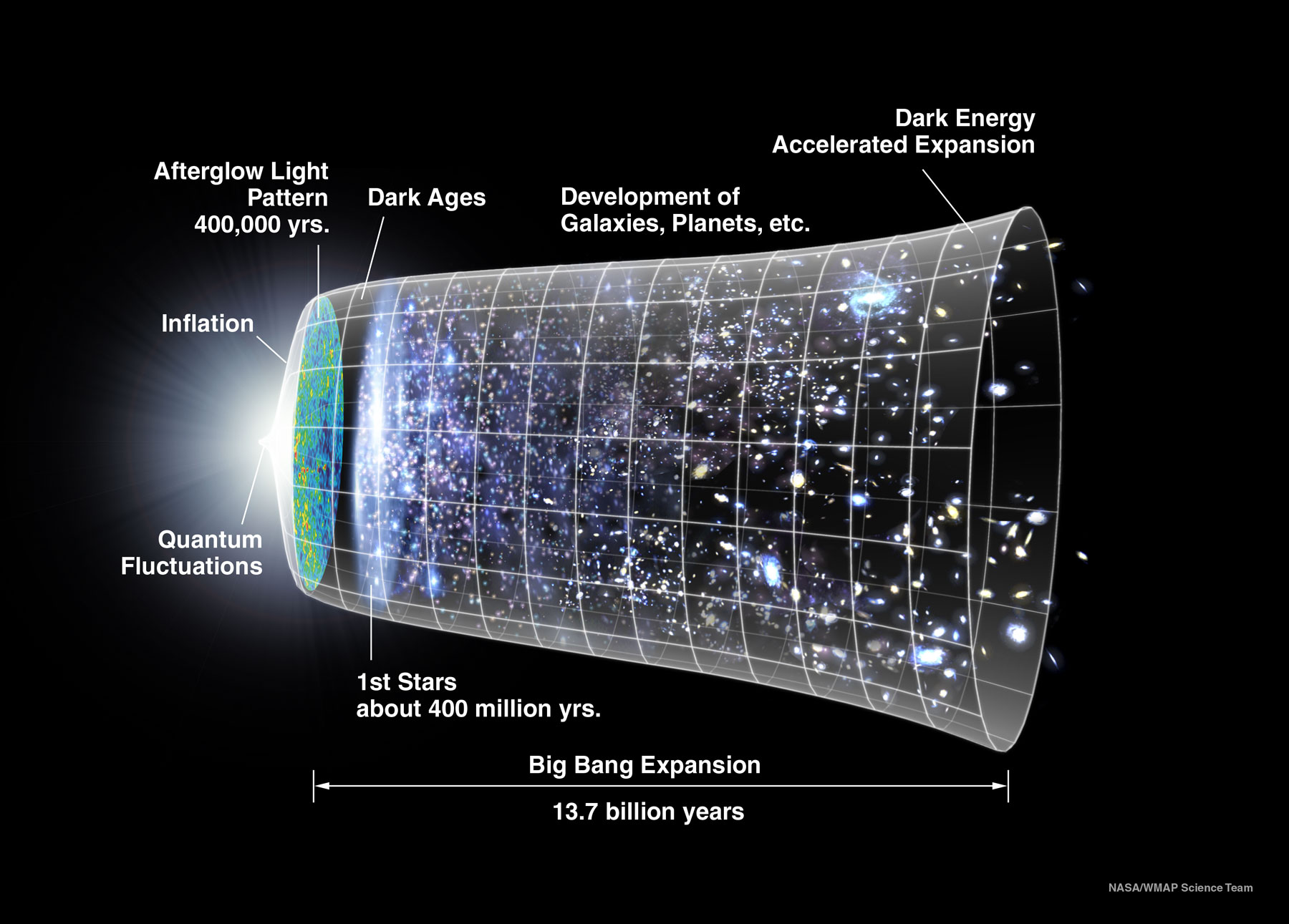}
  {\scriptsize \it Schematic timeline of cosmology, courtesy WMAP/GSFC.}
  \end{center}
\end{minipage}
}

\maketitle

\begin{center}
{\Large \sc Great Surveys of the Universe}
\end{center}



\section{Motivation}

Looking ahead to the next decade and imagining the landscape of
astronomy in 2020, it is clear that astronomical surveys, large and
small, plus extensive follow-up projects, will be a great engine of
progress in our profession.
Surveys have long had a critical role in astronomy, and in the coming
decades will be even more central as we probe deeper into the
Universe.  In fact, one might call the next two decades the {\em ``Era of
Great Surveys''}.  This next generation of surveys will probe a huge
range of astronomical objects and phenomena including planets, stars,
gas, galaxies, background radiation, dark matter, dark energy,
degenerate matter compact objects, black holes, magnetic fields,
cosmic ray particles, neutrinos, gravity waves, and exotica
(particles, topological defects, etc.).

\noindent\underline{\em Why a Great Surveys Program?}
The term ``Great Surveys'' is intended both as a theme
as well as a suite of projects.  The 
inspiration is the ``Great Observatories'' program of NASA and the
key space missions {\it Hubble Space Telescope} ({\sc HST}), {\it Compton
Gamma-Ray Observatory} ({\sc CGRO}), {\it Chandra X-Ray Observatory}
({\sc CXO}), and the {\it Spitzer Space Telescope} ({\sc SST}).  
We are now enjoying the legacy of the Great Observatories in the
tremendous advances that this program has brought to our field.
In addition to the four highly visible centerpieces, we have also
benefited from a broad and balanced portfolio of small, medium,
and large-scale facilities based on ground, sub-orbital, and 
space platforms, with research in diverse areas ranging from
observation and data analysis to modeling and theory.

This case for this decadal review is being framed
in terms of ``central questions'' that are ripe for answering and 
``general areas where there is unusual discovery potential'' that
will define the scientific frontier of the next decade based on
new scientific opportunities and compelling scientific themes.
This Position Paper advocates the overarching theme of a true
Survey of the Universe built up of a diverse range of ``great
surveys'' and the exploitation of these surveys.
A significant number of the proposed decadal activities and facilities
are either explicitly Survey Telescopes or plan to devote significant
amounts of time to survey science.  Others, such as large aperture
narrow field telescopes, are aimed at targeted detailed observations
that are a necessary counterpoint or follow-up to surveys.

\noindent\underline{\em The Great Surveys Workshop:}
The Great Surveys of Astronomy Workshop 
was held 20--22 November 2008
in Santa Fe, NM and was sponsored by the
LANL Institute for Advanced Study and Associated Universities Inc. 
(AUI). This meeting brought together
the community to talk about common issues such as scientific goals,
survey strategy, telescope and information technology, follow-up
observations, coordination of Multi-wavelength resources, theoretical
frameworks, simulations, data management, data products, and other
cross-cutting problems or solutions. To summarize the spirit of the
meeting: ``This meeting will bring together the key workers in the
field to exchange ideas, and identify areas of cooperation,
complementarity, and coordination between surveys and with the greater
astronomical community. The vision is for a future where a variety of
surveys and follow-up programs are underway and flourishing.'' 
Participants spanned wavelength, technique,
research area, and primary funding agency.
The agenda and talks can be found
online at the conference website
\footnote{{\tt http://t8web.lanl.gov/people/salman/grsurveys}}.
There have been other workshops on similar subjects, e.g.
the 2007 KICP Cosmic Cartography 
Conference\footnote{{\tt http://cosmicmaps.uchicago.edu}}.

\noindent\underline{\it Disclaimer:}
The points advanced in this paper are not unique, and
to a large part they are due to discussions with
the participants in the Great Surveys workshop as well as with others
in the field.  I do not do this subject sufficient
justice here, and in particular I do not provide a proper set of 
references as it deserves and refer to only a small number of
papers I know of or have co-authored myself.  
There are a number of superior white papers submitted to the Astro2010 science and
profession panels.  In fact, most the issues presented here are more
fully discussed in a other position papers,
e.g. \cite{WP-Strauss,WP-Ferguson}.  This paper makes use of the more
open process for input to Astro2010 that sets this process 
apart from the previous
ones, and to indulge myself in what can be characterized as a
``manifesto'' of sorts, and as an example of what might be presented
by Astro2010 to those inside and outside astronomy on this topic.

\begin{figure}[t!]
  \begin{center}
  \includegraphics[width=5in]{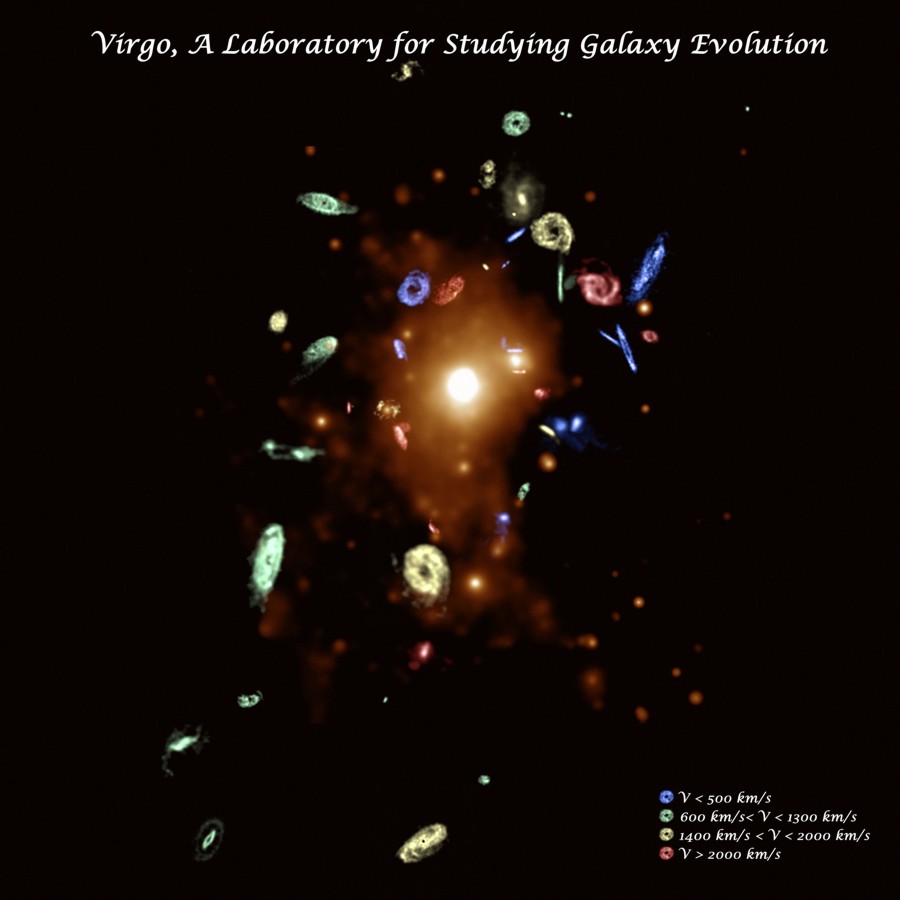}
  \caption{\label{fig:form} 
    A small-scale Survey of the Virgo Cluster of galaxies.
    X-ray image from ROSAT overlaid with VLA HI images of galaxies
    (HI image scale magnified by 10x).  This is but one minor example
    of combining surveys to form a panchromatic view of the sky.
    Image courtesy of NRAO/AUI and Chung et al., Columbia University.
    }
  \end{center}
\end{figure}

\section{Great Survey Science Goals}

The science goals that motivate great surveys are the same goals that
cross-cut all of astronomy and astrophysics.  These are represented 
in the areas defined by the decadal Science Frontier Panels.  See the
excellent Science White Papers that were posted to these panels for
a more complete view of these areas, as well as in reports in the
past inter-decade period (e.g. \cite{q2c,detf}).  
Some survey-related highlights:

\noindent\underline{\em Cosmology and Fundamental Physics:}
Over the past few decades, observational and theoretical advancements
have led to a number of startling conclusions about the physical
nature of our Universe, which are encapsulated in a
``Standard Cosmological Model''.  The cornerstone of this
standard model is that the expansion rate of the Universe is currently
accelerating (requiring an equation of state with negative pressure,
or ``Dark Energy'').  Long-range gravitational forces are dominated by
the Dark Sector: Dark Energy plus effectively collisionless ``Dark
Matter''.

Observationally, the key measurements are of the expansion history
(through its integral, the cosmic distance scale), and growth rate of
structure (which is controlled by the expansion history and the
equation of state).  Large surveys across the electromagnetic spectrum
can be used as precision cosmological parameter probes.  Great surveys
can illuminate dark corners of cosmic time such as during the neutral
era (from recombination to reionization), the radiation dominated era,
the inflationary epoch, and possibly back towards the Planck time.

\noindent\underline{\em Galaxies Across Cosmic Time:}
Galaxies are the ``atoms'' of the large scale structure of the
Universe, aggregated into the molecular Cosmic Web of clusters and
filaments yet retaining identity until merging or dissipating in
encounters. We are now discovering the life cycles of galaxies, and a
frontier for the coming decade is to peer into the mysterious process
of the formation of the first objects.

We also have the opportunity to discover the physiology of galaxies,
and understand the inner workings of the ``sub-atomic'' constituents:
the stars, gas, dust, magnetic fields, high-energy particles, black
holes, and dark matter.

\noindent\underline{\em The Galactic Neighborhood:}
Our corner of the Universe is in many ways a microcosm of the whole,
the understanding of which underpins our view of the macrocosm.  The
Local Group is the archetype of galaxy groups and thus of galaxy
clustering, with the Virgo (e.g. Figure~\ref{fig:form}) 
and Coma clusters our nearest examples of
larger structures.  We now have the possibility of conducting a nearly
complete census of our neighborhood across the electromagnetic
spectrum.  In addition, we can observe phenomena in other nearby
galaxies that have heretofore been seen only in our Milky Way.

Galactic structure studies have been revolutionized by new survey
techniques including microlensing searches, 21-cm neutral hydrogen
mapping, and through astrometry.  Significant advances in these
areas are still to be made through new facilities and surveys.

\noindent\underline{\em Stars and Stellar Evolution:}
Stellar astronomy has also benefited from large surveys, particularly
those targeted at microlensing searches and from the Sloan Digital
Sky Survey (SDSS)\footnote{{\tt http://www.sdss.org}}.  Studies of
the endpoints of stellar evolution --- planetary nebulae, supernova
remnants, white dwarfs, neutron stars, black holes --- have made great
advances.  New phenomena and classes of ``galactic'' objects, such as
Soft Gamma-ray Repeaters and magnetars, have been discovered
based on survey missions and follow-up.

Future surveys will enable the more detailed studies necessary to
build the next generation of theoretical models of stars and stellar
structure.  New phenomena, filling gaps in our mapping of
the timelines of stellar evolution, await discovery.  And finally,
more complete surveys of stars in our own galaxy will shed light on
our own Sun and its future.

\noindent\underline{\em Planetary Systems and Star Formation:}
Charting the sites of star formation in our galaxy and surveying for
planetary systems is a high priority goal for the coming decade.
The recent explosion in exo-planetary research and discoveries 
strengthens the case for the ubiquity of other solar systems and
other Earths.  We are tantalizingly close to being able to truly
search for other life-supporting environments in our Galaxy, and
an ambitious exo-planetary survey and study program is planned for
the coming decade.

Besides being the ultimate genesis of planetary system formation and
thus likely that of life, the process of star formation also greatly
impacts astrophysical systems on all scales, including the
cosmological.  For example, the process of star formation related
``feedback'' into galaxy formation and the intergalactic medium is
a fundamental ingredient in models of the massive clusters of
galaxies that form the backbone of the cosmic web.  Yet truly
predictive star formation models elude us, and are key projects 
for the next decade.  Great surveys provide the observational support
necessary for the construction of these theories.

\noindent\underline{\em Cross-cutting Science:}
In addition to the categories highlighted in the Decadal Survey Science
Frontier Panels, there are cross-cutting areas that drive and derive from
Great Surveys:
\begin{list}{$\bullet$}{
  \setlength{\topsep}{0ex}\setlength{\itemsep}{0ex plus0.2ex}
  \setlength{\parsep}{0.5ex plus0.2ex minus0.1ex}}
\item Things that ``burst in the night'': Transients and Time Domain

      Synoptic surveys on the new generation of telescopes are opening the time
      domain.  This was pioneered for example by gamma-ray burst studies and
      microlensing searches.  In the optical/infrared bands {\sc PanSTARRS} and
      {\sc LSST} are exemplary synoptic survey projects, while the {\sc ATA} at
      radio wavelengths is leading the way.  New capabilities 
      ranging from transient burst and quasi-periodic emission detection,
      variability monitoring, and moving object identification and tracking
      experiments will lead to new and compelling discoveries.

\item Extreme Objects and Physics

      By obtaining extremely large survey samples, studies to find and
      characterize rare phenomena will be enabled.  These in turn will
      test the extremes of physics and astrophysics.  New classes of
      stellar remnants and black holes, and new stages in galaxy, stellar 
      and planetary system evolution, await.

\item New Windows on the Universe

      It is growing harder in astrophysics to open truly new
      ``windows'' in observational space, with most advances pushing
      the current boundaries instead.  However, we are on the verge of
      the start of the first gravitational wave astronomical
      observations.  Due to the highly omni-directional response of
      the detectors, gravity-wave observatories will conduct surveys
      by necessity.  Targets will be teased out of the data stream.
      It may also be that in the coming decade we see the fruition of
      other non-electromagnetic observational experiments such as 
      dark matter searches and more advanced particle and neutrino
      telescopes.

\item New Frontiers and Exploration of the Unknown

      Great Surveys and follow up observations will open frontiers in time,
      distance, and wavelength which will result in new discoveries.
      Although it is difficult to categorize the quest for new
      phenomena in the context of key questions or experiments, the
      exploratory nature of science is at the heart of what
      observational astronomy is all about, and why it is compelling
      to the public as well as to the profession.

\end{list}

\section{Realizing Great Surveys}

It will take more than a position paper to realize a Great Surveys
program and to make this theme more than cosmetic.  There are many
challenges facing astronomy and astrophysics in the coming decades.
Though these are framed in terms of surveys, they have more general
significance, and will be found in any number of Decadal white papers,
position papers, and project proposals.  In particular, see 
\cite{WP-Ferguson} for more on data processing and analysis issues.
I do not claim any particular insight here!

\subsection{Ingredients in a Great Surveys Program}

\noindent\underline{\em Survey Infrastructure:}
The hardware and software needed to carry out and process a large
survey is substantial, and particularly on the information technology
side is becoming increasingly complex.  Large surveys must be
well-planned and well-managed.  Even though much of the value comes
from use of the published products by persons outside the original
collaboration (and is thus off the ledger), the public survey approach
necessarily adds requirements on the survey sophistication, for
example in tight quality control.  This must be designed and built
in from the start.

\noindent\underline{\em Data Products:}
The ultimate payoff in survey science is in the form of
the raw (e.g. time-ordered) data and the processed survey Data
Products such as images and object catalogs.  This is what
the general astronomical community will be expected to use.
Assurance of the quality of these products is paramount
(see below).  Modern archives and data releases for programs
and missions such as {\sc SDSS}, {\sc HST}, {\sc CXO}, and
{\sc SST} have led the way in setting the standards for what
products users should have access to.

\noindent\underline{\em Data Tools:}
The utility of publicly accessible data products is also dependent
on the quality and usability of tools that can turn the data into
science.  Most areas of astronomy have one or more standard data
reduction packages or suites of applications that make use of 
generic environments (e.g. Python, IDL, Matlab, Mathematica).  
Significant effort is going into next generation data reduction,
visualization, access and analysis tools within projects.  In 
many cases common standards and multi-use software packages would
improve the ability of casual multi-wavelength astronomical users to 
make use of survey data.  National initiatives facilitating the
development of cross-cutting tools be an effective way to unify
common efforts, though this should be balanced against scope creep
and added complexity that often occurs with large software projects.

\noindent\underline{\em Universal Access:}
The ultimate payoff in survey science comes from enabling astronomers
that would not otherwise be able to use and contribute as part of
the survey team.  Publicly available data on a short
but reasonable timescale after observation is the necessary
first step.  SDSS was spectacularly successful in tapping the 
wider community for value-added science beyond that of the 
core survey group.  Future surveys have even greater potential
for nearly universal astronomical utility.  Once
data is public, there needs to be a way for astronomers to use it.
The establishment of the national and international Virtual Observatory 
(VO) projects\footnote{{\tt http://www.us-vo.org}}
this past decade was a step towards enabling the
widest possible use of data resources.  For Great Surveys, VO 
applications will need to have even wider application, yet have
much deeper visibility into the actual survey data in order to
maximize the quality and efficiency of its performance.

\noindent\underline{\em Opportunity for Innovation:}
It may not be apparent that surveys are engines for innovation, but
the success of a Great Surveys program will hinge on our ability to
spur and in turn incorporate novel developments by the community.
Unlike first generation surveys based on physical media (plates and
prints), such as the pioneering Palomar Observatory Sky Survey (POSS), 
and even early digital surveys and catalogs, 
modern sky surveys such as SDSS have
reached a level of data richness and sophistication where new
approaches to data mining and analysis can have a huge impact on
survey utility.  The extremely large data rates and volumes from
future surveys will necessitate advances in processing and analysis,
both within the survey pipeline infrastructure and in user-oriented
post-processing.  It will be critical to invite this opportunity and
to build the survey to maximize its richness and to facilitate
innovation.

\noindent\underline{\em Great Surveys Need Great Theories:}
The interplay between theory, modeling, observation, and 
data analysis is what allows us to make scientific progress
in a classical sense.  For example, in many areas we have
been able to build a paradigm or ``standard model'' that is a
synthesis of the knowledge in the field.
Advances in our capability to carry out detailed simulations
of astrophysical processes and cosmological volumes has
revolutionized the field.  The products of these simulations
are datasets in their own right, and many observers and theorists
have made use of these resources.

In the coming decade, progress must be made in a number of areas of
theory and modeling. This case is made more directly in a number
of other position papers and in numerous science white papers,
e.g. \cite{WP-Myers,WP-Kravtsov,WP-Rudnick} to cite only a few.
Key issues include:
\begin{list}{$\bullet$}{
  \setlength{\topsep}{0ex}\setlength{\itemsep}{0ex plus0.2ex}
  \setlength{\parsep}{0.5ex plus0.2ex minus0.1ex}}
\item Analytic and Semi-Analytic Modeling for Precision Cosmology

      To a large extent, the measure of our ``understanding'' of
      astrophysics and cosmology lies in our ability to 
      construct analytic theories of the Universe that capture 
      key aspects of the physics that can apply to the observations.
      In more phenomenological cases, such as deriving observable
      functions (e.g. mass functions, luminosity functions, population
      counts) semi-analytic models can capture the salient features
      of the theory.  A vibrant and adequately supported theoretical
      community is required for a Great Surveys (or any other) program
      to have meaning.

\item Next Generation Models and Simulations

      The process of extraction of model parameters from large surveys
      will require advanced simulation and modeling technologies to 
      be effective.  For example, the recent renaissance of cosmology
      and the progress towards precision cosmology has been sustained
      by innovations in methodology for the extraction of cosmological
      parameters from observations of the cosmic microwave background
      and large-scale structure in comparison with fast Boltzmann
      codes and N-body simulations.  Increases in survey size coupled
      with tighter precision requirements for the next-generation of
      surveys will need simulations with larger numbers of cells or
      particles, greater accuracy, with greater efficiency.  Besides
      cosmological simulations, the modeling of galaxy formation, 
      star formation and stellar populations, and the genesis and
      properties of planetary systems are key areas for advance.

\item Tools and Access

      As with the processing of the survey data itself, the modeling
      and simulation of next generation surveys requires specialized
      tools and scientists to create and use them.  Besides widely
      used standard codes (e.g. {\sc Gadget} \cite{gadget}), 
      there is a healthy 
      number of simulators who write their own codes, keeping the
      pace of innovation moving forward.  On the other hand, some 
      level of interoperability and intercomparison between codes
      is required for quality assurance (e.g. \cite{cccp}).

\end{list}

\subsection{Great Survey Challenges}

In order to get maximum science return from our investment in
facilities, science programs, and people, there are a number
of cross-cutting issues that need to be addressed.  See
\cite{WP-Strauss} for more detailed discussion.
These topics include:
\begin{list}{$\bullet$}{
  \setlength{\topsep}{0ex}\setlength{\itemsep}{0ex plus0.2ex}
  \setlength{\parsep}{0.5ex plus0.2ex minus0.1ex}}
\item Data Management and Optimal Processing for Large Surveys

      One of the salient characteristics of the next generation
      Great Surveys is the tremendous data rates and volumes 
      expected. Handling this data in the pipeline and early
      processing stages will be an extreme challenge.

\item Accessibility and Distribution

      Getting the rawer formats of the data out to the public
      is also difficult given the huge volumes.  This will 
      likely require partnerships with high-performance 
      computing and data centers, as well as widely accessible
      VO tools.

\item Data Mining and Science Extraction

      Survey end-users (both internal and external) will need to 
      dig through the data and extract model parameter constraints,
      calculate object counts and classifications, and measure
      statistics and/or detailed properties on data products.
      Again, astronomical data volumes present a challenge here,
      particularly for use by astronomers who might not have easy
      access to the top computational facilities.

\item Interfacing Simulations/Models with Survey Data

      As mentioned above, big surveys tend to require big simulations
      and extensive modeling, thus this aspect ends up inheriting
      the same issues as the data side.

\item Common and Transportable Simulations

      Again, as pointed out previously, some level of intercomparison
      between simulations is needed, as well as a mechanism to
      exchange data and simulation results.

\item Data Product Quality Assurance

      As both surveys and simulations become large and expensive, 
      it becomes paramount to monitor and assure the quality of the
      data, its products, and the results of analysis.  You may only
      be able to run the pipeline or simulation a few times (or once
      in pseudo-realtime) and thus there will be diminishing space 
      for error tolerance.  This is likely to be a non-trivial 
      challenge for future surveys and modeling.

\item Survey Strategies and Maximizing Survey Complementarity

      There are many proposed survey and follow-up programs proposed
      for the coming decade.  Although an attitude of ``build the
      survey and users will come'' has some merit, truly leveraging
      the diverse avenues of exploration into a reasonably coherent
      Great Surveys program will require some minimal level of 
      coordination and exchange of ideas.  Perhaps this can be done
      through informal workshops like the Great Surveys meeting, or
      perhaps more formal mechanisms are needed.

\item Multi-wavelength Follow-up Needs of the Great Surveys

      For every survey telescope, there needs to be some reasonable
      number of facilities capable of follow-up observations, e.g.
      {\sc Chandra}, Gemini, HST, Keck, Magellan, VLA and VLT
      (to name only a few) for SDSS.  Adequate balance in the funding
      between surveys and 
      exploitation is required.  Generally, different telescopes are
      used for survey and follow-up, but in some cases (e.g. the SKA)
      the {\em same} instrument will do both the survey and exploitation,
      complicating the operational model for that facility.

\end{list}

\subsection{Survey Sociology: It Takes a Community}

An ambitious endeavor such as a Great Surveys program cannot happen
in a vacuum.  It is more than the sum of its telescopes, technology,
and even its data products. Its success is primarily driven by the
people that develop the design, build the infrastructure, create
the technology, operate the survey, process and publish the data, use the
products, turn it into science, and communicate the import and context
to the public and to the next generation of young astronomers.

One of the surprises that I took away from the Great Surveys workshop
was the extent of the passionate discussion of issues that had to
do with the sociological and professional aspects of survey prosecution and
use, rather than the more technical tel scope and data problems.  There
were concerns about the perception of ``survey scientists'' in the
wider and more traditional astronomy community, for the
employment future of students and post-docs who work on survey
programs, and for the assignment of credit in publications for
contributions to surveys.

Some key issues:

\noindent\underline{\em Survey Science \& Sociology:}
Surveys are going to play an even larger part of future astronomy than
they do now, and the survey projects as well as our profession as a
whole must come to grips with the sociological implications of this
quiet revolution.  There must be room for small groups and individuals
as well as large groups to play visible and vital roles in surveys.
Student education clearly must be part of Great Surveys (as it is
now), and there must be viable and interesting career paths for the
students and professionals involved at all levels of survey work.
Evolution of surveys from small teams to large collaborations is
inevitable, and we must come to grips with this.  

In addition, the advent of large survey projects and the availability
of accessible survey data products has led to the appearance of new
kinds of astronomers --- ``data wranglers'', survey scientists,
analysis experts, software astronomers, etc. --- that were a very
small minority even 15 years ago.  Since the proof is in the science,
the acceptance of these new facets of the profession has been driven
by the need and utility, but a perceived lack professional recognition
still troubles the field.

\noindent\underline{\em Great Surveys Great Astronomy:}
It cannot be stressed too greatly that the ultimate goal in everything
we do is to produce the highest quality, highest impact, most
innovative, and most useful science possible with our people,
projects, and facilities.  It is for Great Astronomy that we do 
Great Surveys.

A relatively new aspect of surveys (in the SDSS age, as compared to
the POSS era) is the development of efficient and usable mechanisms
for the distribution of survey products and results to the astronomy
community and to the public.  The VO network was established to 
enable this on the widest scale possible, and one of the challenges
of the coming decade will be to make the virtual observatory concept
truly work or to find a different better paradigm.  At the Great
Surveys workshop there was some discussion on the approaches the
surveys, super-computer and data centers, and VO might take for better
partnerships and improvement.

Making the fruits of Great Surveys visible to the greater scientific
community, to the public, to funding agencies, to politicians, and to
other stakeholders is also of critical importance.  This has always
been a challenge, often requiring a different set of communication 
skill than those possessed by most astronomers and astrophysicists.
The increasing emphasis on EPO in the profession highlights the need
for increased support here.  We are doing relatively well (compared to
other basic science and mathematics fields, in my opinion) but there
is clearly room and requirement for improvement.

\noindent\underline{\em Not Everything To Everyone:}
In this position paper, I advocate an umbrella of Great Surveys as a
focal point for Astro2010.  But clearly not everything is a survey,
nor is everything an astrophysics experiment.  There must be balance 
in our national astronomy program, and diversity at all levels should
be encouraged.  Even if the Great Surveys banner is dismissed, surveys
will clearly be a large part of the picture and the issues highlighted
here will need to be grappled with.

One of the most interesting discussions in the past few years on the
sociology of the field was instigated by a paper
on the ``conflict'' between experiment style astrophysics and more
general observatory driven astronomy \cite{white07}.  This discussion
is even more relevant in the coming decade, as it will be challenging
to maintain diversity of our program in the face of increasing project
scope and cost and larger collaborations and project teams. 

\section{Conclusions}

In this position paper, the concept of a Great Surveys meta-program
was advanced, along with the presentation of a number of issues and
challenges that astronomers face in the coming decades.  These are
not new to the field (some even predate SDSS --- gasp!) and are 
certainly better described in other white papers.  The evolution
of astronomy and astrophysics will continue regardless, and the
growth of projects means that key programs and projects will continue
to span decades in scale and scope.
The Astro2010 panel has their work cut out for them to reconcile the
diverse needs and wants of the community and to synthesize a
compelling and coherent vision for the future!


\noindent\underline{\it Acknowledgments:}
STM acknowledges the support of the Orson Anderson Scholarship of
the IGPP while on sabbatical at the Los Alamos National Laboratory,
during which time the concept of a Great Surveys program was fleshed out.
The generous support of the LANL Institute for Advanced Study, and of
Associated Universities Inc. (AUI) allowed the Great Surveys Workshop
to take place.  The National Radio Astronomy Observatory
is a facility of the National Science Foundation operated under cooperative
agreement by Associated Universities, Inc.

\noindent\underline{\it Another Disclaimer:}
The issues raised in this position paper owe a great deal to the
excellent talks presented at the Great Surveys workshop, and to
discussions between the participants.  I apologize for not providing
a proper conference summary with direct talk references 
(see the conference website).  The views expressed here are my own and
do not necessarily reflect those of my employer (NRAO) nor of the
meeting participants.

\end{document}